\documentclass[10pt, twocolumn]{article} 
\usepackage{graphicx}
\usepackage{amssymb}
\usepackage{authblk}

\begin{document}

	\title{Images from the Mind: BCI image evolution based on RSVP of polygon primitives}
  \author[1,2,3]{L. F. Seoane}
  \author[3]{S. Gabler}
  \author[3,4,5]{B. Blankertz}
  \affil[1]{ICREA-Complex Systems Lab, Universitat Pompeu Fabra, Dr. Aiguader 80, 08003 Barcelona, Spain. }
  \affil[2]{Institut de Biologia Evolutiva, CSIC-UPF, Psg Barceloneta, Barcelona Spain. }
  \affil[3]{Bernstein Center for Computational Neuroscience, Technische Universit\"at Berlin, Germany. }
  \affil[4]{Neurotechnology Group, Berlin Institute of Technology, Berlin, Germany. }
  \affil[5]{Berstein Focus: Neurotechnology, Berlin, Germany. }

	\date{\today}

	\maketitle 
	
	\begin{abstract}

    This paper provides a proof of concept for an EEG-based reconstruction of a visual image which
is on a user's mind. Our approach is based on the Rapid Serial Visual Presentation (RSVP) of polygon
primitives and Brain-Computer Interface (BCI) technology. The presentation of polygons that
contribute to build a target image (because they match the shape and/or color of the target) trigger
attention-related EEG patterns. Accordingly, these target primitives can be determined using BCI
classification of Event-Related Potentials (ERPs). They are then accumulated in the display until a
satisfactory reconstruction is reached. Selection steps have an average classification accuracy of
$75\%$. $25\%$ of the images could be reconstructed completely, while more than $65\%$ of the
available visual details could be captured on average. Most of the misclassifications were not
misinterpretations of the BCI concerning users' intent; rather, users tried to select polygons that
were different than what was intended by the experimenters. Open problems and alternatives to
develop a practical BCI-based image reconstruction application are discussed.

  \end{abstract}

	\section{Introduction}
		\label{sec:1}

    We introduce a Brain-Computer Interface (BCI) application based on Rapid Serial Visual
Presentation (RSVP) of polygon primitives for image reconstruction. Our paradigm relies on the
decomposition of a collection of images (figure 1) into a set of approximate constituent parts
(polygon primitives). These primitives are presented to the experimental subjects in bursts. Event
Related Potentials (ERPs) associated with the oddball paradigm \cite{PatelAzzam2005, Polich2007} are
used as EEG correlates to detect those pieces that contribute to the reconstruction of the target
image (which is on the user's mind) in an incremental fashion. The operational basis (RSVP) has
already been employed for BCI-based gaze-independent spellers \cite{AcqualagnaBlankertz2010,
AcqualagnaBlankertz2013} and an Icon Messenger \cite{AhaniErdogmus2014}. The results presented in
this paper invite us to be optimistic about this new paradigm for BCI image reconstruction. More
experiments should be carried out to expand the design outside the chosen collection of images,
ideally moving towards a free-painting device. Some ideas are proposed based on the outcome of the
current work. The experiments also suggest ways to increase the reliability, speed, and accuracy of
the current framework.

    Our work draws from recent advances in Cortically-Coupled Computer Vision (C3Vision)
\cite{GersonSajda2006, ParraSajda2008, SajdaChang2010} where human vision is enhanced through an
efficient data mining of EEG patterns while stimuli are presented. Such an approach has shown to
boost the search for target images, and how to speed up the localization of salient details from
within a large image \cite{BigdelyMakeig2008, HuangHild2010}. Even further improvement can be
achieved with a {\em closed loop} philosophy in which human vision and EEG are not only coupled, but
engage in a cycle where the artificial intelligence behind the EEG-based classifiers offers feedback
in real time \cite{PohlmeyerSajda2011, UscumlicMillan2013}. These seminal studies provide
interesting insights about the capabilities of such BCIs and suggest that the limits of our own
design could be pushed further.\\ 

    In \cite{BasaLee2006} we find an early approach closer to the line of research that we pursue.
Physiological states -- including EEG and indicators such as ventilation, heart rate, and others --
are correlated to `positive' and `negative' feelings elicited by visual stimuli. The ability to
prompt positive reactions is then used as fitness functions for new, randomly generated images that
evolve by means of a genetic algorithm. The decisions based on these cues are compared offline to
the deliberate choices made by the experimental subjects who should decide what images were more
artsy. These physiological correlates predict up to $61\%$ of the subjects’ choices. It is more
complicated to assess the global goal of this research -- i.e. evolving art.

    The main drawback of this paradigm is the extended exposure time needed to gather the relevant
physiological data. Subjects are exposed to pictures for one second so that neural correlates of
mood can be recorded. This stretches a single generation of the genetic algorithm to $12$ minutes.
Because of this large exposure, it is likely that only coarse features of a picture are relevant for
selection. This might be a problem for practical applications, if we wanted to direct the evolution
towards detailed visual features. Additionally, the only selective pressures are ``positive''
feelings elicited by the drawings, for which it is difficult to quantify a progress: Does the last
image render more positive feelings than the original one? How long can we advance in positiveness
before the physiological correlates saturate? These thorny details, rather than flaws in the
original design, reveal the difficulty of the task under consideration.

    While the authors of \cite{BasaLee2006} intended to explore the undetermined space of art and
feelings evoked by drawings, Shamlo and Makeig \cite{ShamloMakeig2009} sketch a procedure to evolve
towards a definite target image. Bursts of randomly generated drawings are presented to the
subjects, whose EEGs are recorded. A burst might or might not contain a target image that resembles
``two eyes''. The subject presses a button after each burst to indicate if the target has been
consciously spotted out. EEG patterns are extracted that correlate with the presence of the target
image during a burst. The authors focus on processing the data during a posterior offline analysis.
Using ten fold cross-validation on data from a single trial, very high accuracy is obtained (up to
$0.98$ area under ROC curve of correctly classified target pictures).

    Encouraged by these good results, and in the spirit of C3Vision, the authors propose to use
their BCI paradigm to bypass the slow feedback that subjects have to provide manually nowadays. One
straightforward application suggested is to use this EEG activity to evolve images. Currently
existing software \cite{picBreeder}, accepts user-provided images and attempts to evolve them
towards a desired target picture. Random mutations and crossover are applied to the original seed to
generate new drawings. In the standard approach, the user manually chooses among the new candidates
that fall closer to an arbitrary goal (e.g. ``two eyes'' as in \cite{ShamloMakeig2009}). This manual
procedure is cumbersome. Using EEG correlates to identify what images should be fed to the algorithm
could speed up the evolutionary process. Unfortunately, such interesting possibility is only
suggested in \cite{ShamloMakeig2009} based on the exceptional performance of the EEG-driven
classification task. Actual experiments should be implemented to test the complete 
BCI--evolutionary-algorithm loop. When this is done, it will be possible to address an important
aspect of the BCI design. The performance reported in \cite{ShamloMakeig2009} refers to the
identification of target images within bursts that the experimental subject {\em manually}
identified as containing a target. The high accuracy of this manual selection ($94\%$ correct
classifications) indicates that the task (the identification of a broad feature such as ``two
eyes'') might be simple enough. It is an open question about how accurate the procedure will be once
the tasks become harder, e.g. if we would attempt to evolve more detailed visual structures.

    The {\em P300-Brain Painting BCI} \cite{KublerHosle2008, HalderKubler2009, MunssingerKubler2014}
inspired by the early P300-Speller BCIs \cite{FarwellDonchin1988} is the device closer to ours in
goals and performance. Using the oddball paradigm, colors, shapes, and a variety of tools are
selected from a matrix of highlighted rows and columns. The selected operators modify an existing
canvas in a similar way that a mouse interface would do. In \cite{MunssingerKubler2014} this BCI was
evaluated in healthy and ALS patients finding, for certain design specifications, performances
comparable to those of the equivalent matrix speller. The insights from these works -- in terms of
speed and accuracy -- are complicated to translate to ours due to the important differences in
implementation. We investigate two different -- albeit similar in purpose -- designs, and this will
complicate the comparison between BCIs as discussed below.

    The painting tools available in \cite{MunssingerKubler2014} are rather scarce: two shapes
(square and circle), eight colors, and four sizes. Despite this potential limitation, the subjects
can produce quite complex paintings (figure 3 in \cite{MunssingerKubler2014}). If we wanted to
incorporate more shapes or colors, we might face an important limitation since the number of
painting tools must fit in the symbol matrix: a larger matrix (more tools) will translate in lower
selection accuracy. As the authors point out, this problem affects the movement of the mouse cursor
too: only eight directions are allowed along which the cursor moves one single unit at a time. It is
suggested to use sensorimotor rhythms to control a mouse pointer over the canvas. That would release
space in the symbol matrix for other needs. Despite these minor issues, the BCI in
\cite{MunssingerKubler2014} must be regarded as an important breakthrough for BCI-painting.

    The paradigm explored in \cite{BasaLee2006} relies on physiological signals. This slows down the
interface drastically, as tens of seconds are necessary to collect reliable data. But the approach
in \cite{ShamloMakeig2009} and others \cite{PohlmeyerSajda2011, UscumlicMillan2013} set an
interesting precedent for our BCI given the high rate at which stimuli are presented (up to $12$
Hz). We chose more conservative rates (around $3$ Hz) for the proof of concept introduced here.
Another crucial difference between most examples in the literature and ours is that we proceed
bottom-up to reconstruct a set of images: our stimuli are polygon primitives that might resemble
smaller details of a target drawing, which allows for a finer grained reconstruction than those in
\cite{BasaLee2006, ShamloMakeig2009}. In these studies, each stimulus is a whole picture and targets
are based on broader features such as the existence of two eyes. This would be important, if we
wanted to extrapolate the technical setup (mainly the stimulus rate) to our design. Furthermore
experiments actual image reconstructions were undertaken, thus we offer the first serious test of
this novel and promising BCI paradigm.\\

    The paper is organized as follows: In section \ref{sec:2} the proposed BCI is described along
with the experimental and data analysis details. In section \ref{sec:3} the results are summarized.
We close with a discussion of these results (in comparison with the existing literature) in section
\ref{sec:4}, where future lines of research are proposed together with possible improvements and
alternatives to the current design. Appendix \ref{app:1} includes a description of the choices made
regarding the preprocessing of the images and the experimental setup. This is compared to similar
BCI schemes that inspired the research. Appendix \ref{app:2} analyzes in detail an anomaly found in
some image reconstructions.

	\section{Methods}
		\label{sec:2}

		\subsection{Participants}
			\label{sec:2.01}

      Ten participants (five women and five men, ages ranging from early twenties to early thirties)
took part in the experiment on a voluntary, non rewarded basis. One of the subjects was associated
to the BCI research group. This subject and two others had previous experiences with BCI
experiments. The rest of the subjects were naïve with respect to BCI technologies. All of them had
normal or corrected to normal vision and did not report any health issues during the experiment.

		\subsection{Apparatus}
			\label{sec:2.02}

      EEG was recorded at 1000 Hz using BrainAmp amplifiers and an ActiCap active electrode system
with $63$ channels (Brain Products, Munich, Germany). The electrodes used were Fp1,2, AF3,4,7,8, Fz,
F1-10, FCz, FC1-6, FT7,8, Cz, C1-6, T7,8, CPz, CP1-6, TP7,8, Pz, P1-10, POz, PO3,4,7,8, Oz, O1,2.
All electrodes were referenced to the left mastoid using a forehead ground. For offline analyses,
electrodes were re-referenced to linked mastoids. All the impedances were kept below $10 \>
k\Omega$.

      Stimuli were presented on a $24$'' TFT screen with a refresh rate of $60 \> Hz$ and a
resolution of $1920 \times 1200 \> px^2$. The experiment was implemented in Python using the 
open-source BCI framework Pyff \cite{Pyff} with Pygame \cite{pygame} and Vision-Egg
\cite{visionEgg}. Data analysis and classification were performed with MATLAB (The MatlabWorks,
Natick, MA, USA) using an in-house BCI toolbox (www.bbci.de/toolbox).

		\subsection{Design and procedure}
			\label{sec:2.03}

      The design of the experiment includes a pre-processing of the images to extract the primitives
that are shown during RSVP bursts. Important choices regarding stimulus presentation were also made.
Because we explore a novel paradigm, almost any design feature is open to debate. In the following,
we report the actual choices made for the experiment. For an introduction and discussion of the
other possibilities the reader is referred to appendix \ref{app:1}.

			\subsubsection{Preprocessing of target images}
        \label{sec:2.03.01}

\begin{figure}[htbp]
          \begin{center}
            \includegraphics[width=0.5\textwidth]{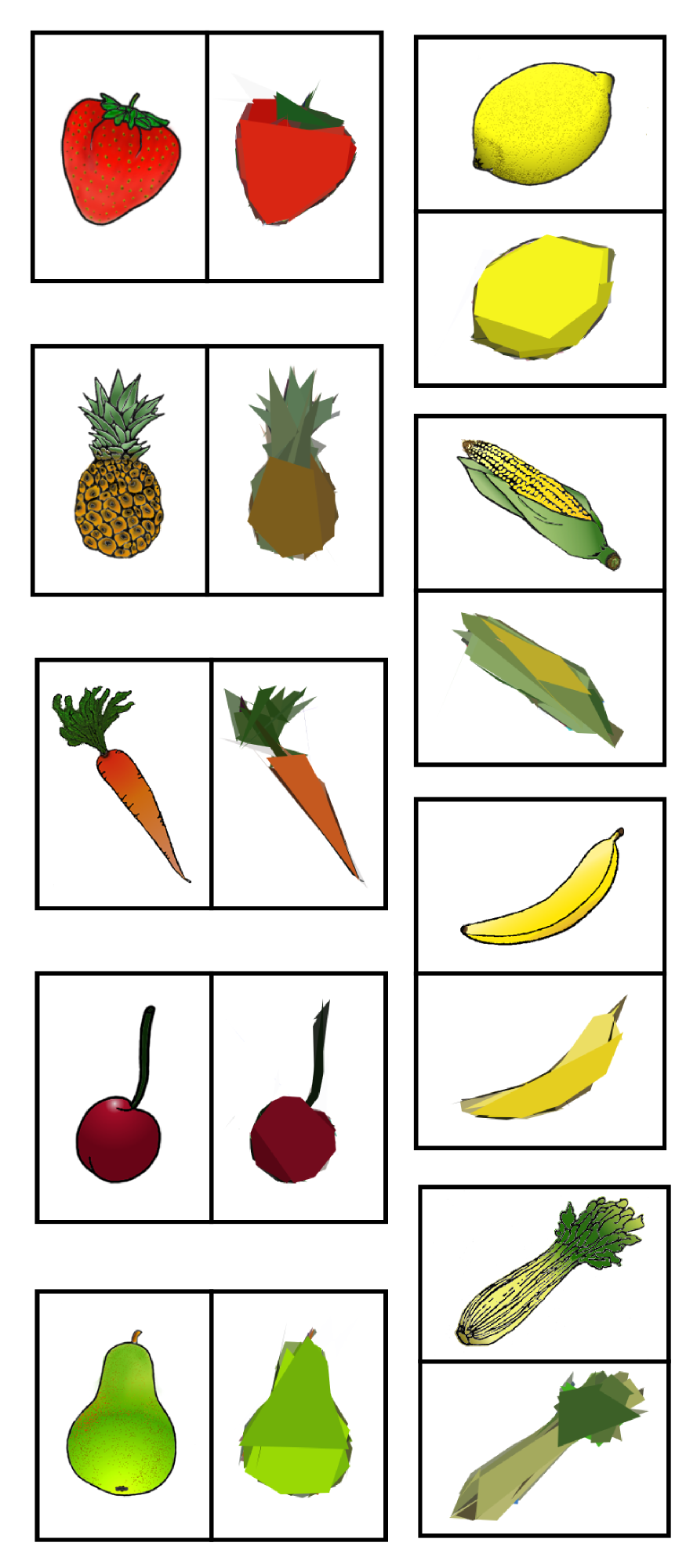}

            \caption{{\bf Images selected as BCI reconstruction targets}. The chosen images comprise
very basic and recognizable shapes with bright colors. Besides, some of the drawings present
relatively small details that become important as the reconstruction advances. We also show the
polygon decompositions obtained for each image. }

            \label{fig:1}
          \end{center}
        \end{figure}

        The nine drawings of fruits and vegetables collected in figure \ref{fig:1} were chosen as
potential targets for reconstruction from the revised Snodgrass and Vanderwart's object database
\cite{RossionPourtois2001}. They are easily recognizable, have bright, plain colors, and combine
basic shapes with some minor details -- such as the stalk of a cherry -- that could act as landmarks
during a reconstruction task. They are always displayed upon a blank background.

        Similar to the role played by characters in RSVP spellers \cite{AcqualagnaBlankertz2010,
AcqualagnaBlankertz2013}, we need nuclear units that constitute stimuli during the bursts of our BCI
paradigm. These stimuli must be able to reconstruct the target images as they accumulate in the
screen. We submitted the chosen images to a preprocessing step during which a genetic algorithm
\cite{genAlgorithm} extracted a series of primitives (polygons) that approximated the drawings up to
a satisfactory degree (final product shown in figure \ref{fig:1}). We refer to these primitives
throughout the text as the {\em polygon decomposition} of each of the targets. A brief description
of the genetic algorithm is found in appendix \ref{app:1.01}. The actual code that we used and the
values of different parameters chosen for our implementation can be found in a public repository
\cite{codeGA}.

        The resulting decompositions contain between $4$ and $10$ polygons (with $6.7$ on average)
depending on the target drawing. We restricted polygons to have between $3$ and $7$ vertices and
solid colors -- i.e. no transparency was allowed. Polygons are overlaid on a blank background and on
each other, thus their order matters for the reconstruction task.

        The {\em fitness} function used as a selection criterion by the genetic algorithm (see
appendix \ref{app:1.01}) is a pixel-by-pixel distance between an image and its polygon
decomposition. By computing the fitness drop of a decomposition when one polygon is removed, we have
a measure of the impact of each polygon in the final reconstruction. We dub this measure {\em visual
information} and report it as a percentage. When running the BCI experiments we only used polygons
with a visual information higher than $3\%$ ($15\%$ during the calibration phase -- see sec.
\ref{sec:2.03.02}). On average, the polygons eventually selected for the experiments bore a visual
information of $13.5\%$ with a standard deviation of $13.4\%$.

        This visual information also allows us to rank the polygons within a decomposition according
to their importance in the reconstruction. We used this ranking to select what target polygons are
displayed in the initial bursts of the reconstruction, and we moved towards less informative
polygons as the reconstruction proceeded, as explained in section \ref{sec:2.03.02}.

        Finally, the oddball paradigm targets are displayed among allegedly neutral non-target
stimuli. Whenever a picture was selected for reconstruction, the polygons in its decomposition were
considered target primitives. The polygons from the decomposition of all non-target images were held
in a pool from which non-target primitives were drawn at the beginning of each burst (see appendix
\ref{app:1.01} for discussion).

			\subsubsection{Experimental setup}
				\label{sec:2.03.02}

\begin{figure*}[htbp]
          \begin{center}
            \includegraphics[width = 15 cm]{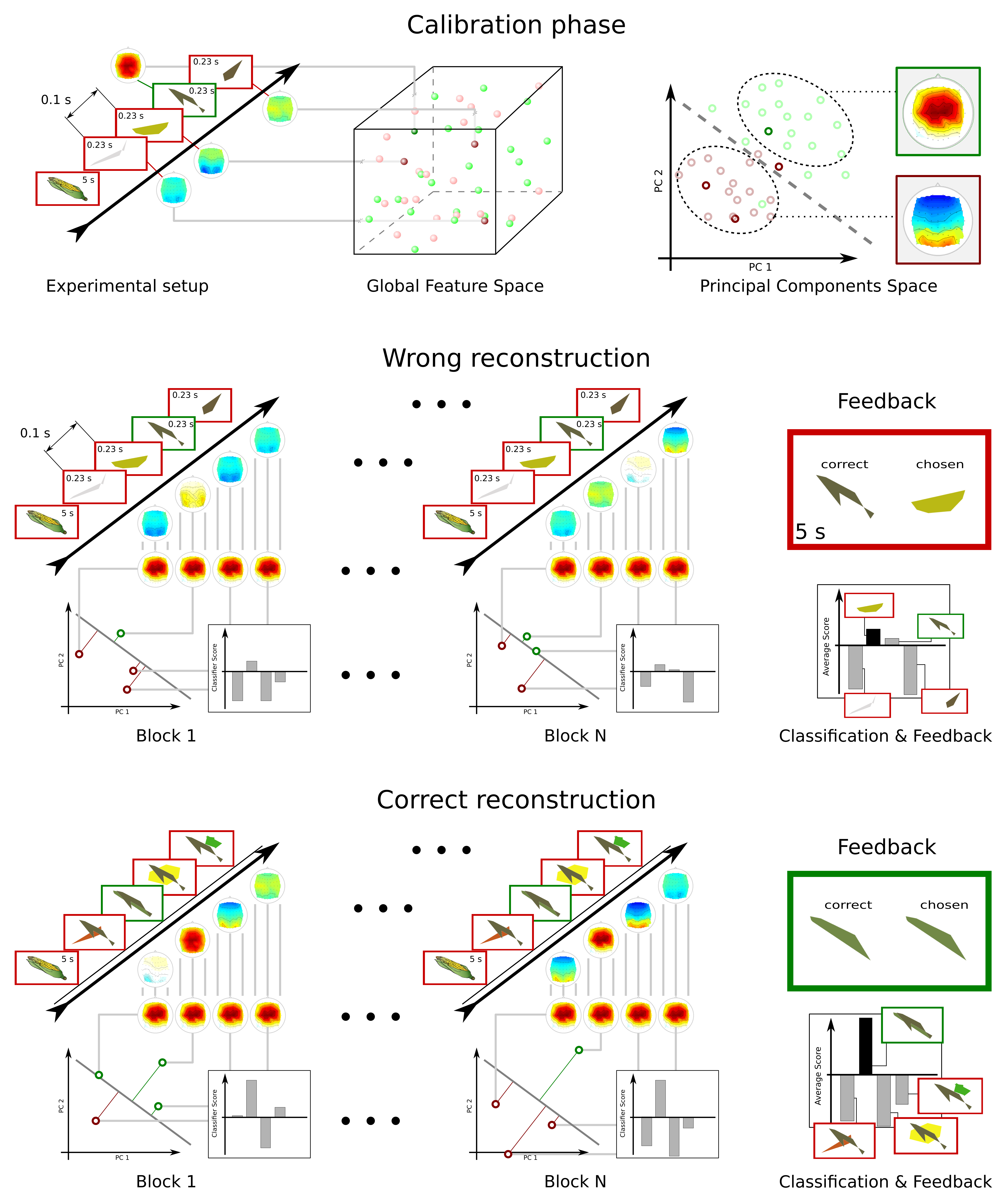}

            \caption{{\bf Schematic representation of bursts during different phases of the
experiment}. {\bf a} {\em Calibration phase}: Bursts of polygons are presented to the subjects with
no feedback between bursts. A target picture is randomly chosen for each burst. Stimuli in this
phase are restricted to polygons bearing more than $15\%$ visual information to ensure that ERPs are
reliably elicited and measured. {\bf b} {\em Reconstruction phase}: A target picture is chosen and
remains target for several bursts until its reconstruction is completed. Each burst presents new
target polygons upon the background where previous target primitives accumulate, thus implementing
the image reconstruction. After each burst a feedback is provided showing the correct polygon and
the polygon selected by the classifier. Despite the classification result, the correct polygon is
always kept in successive bursts so that the reconstruction can proceed. An example of the
reconstruction implemented is shown in \cite{videoCherry}.}

            \label{fig:2}
          \end{center}
        \end{figure*}

        The experiment consisted of a {\em calibration} phase and a {\em reconstruction} phase,
schematically depicted in figure \ref{fig:2}. In either phase, a drawing from figure \ref{fig:1} was
selected as a target and displayed for $5$ seconds prior to each burst. A burst consisted of the
rapid presentation of polygons that included target and non-target stimuli -- i.e. primitives from
the decomposition of the target and non-target images respectively. For each burst, $5$ non-target
polygons were selected for display together with the (one) corresponding target polygon.

        A burst involves a number of blocks that may be different for the calibration and
reconstruction phases. Each block consists of the display of all $6$ polygons in a shuffled order.
Each polygon is shown just once per block. The only constraint to the random order within a burst is
that no primitive may appear twice in immediate succession. Because a burst is a series of blocks,
this restriction only affects the randomness of a block given the last stimulus of the previous
block. The stimulus onset asynchrony (SOA) between successive polygons was $330 \> ms$: during $230
\> ms$ the visual stimulus (the polygon) was shown and during the last $100 \> ms$ the corresponding
polygon was substituted by a transparent rectangle -- i.e. by a {\em void} stimulus.

        {\em Calibration phase}: We restricted the visual stimuli to those polygons contributing
more than $15\%$ visual information. By avoiding polygons that might not be reliably recognized as
targets by the participant (due to little visual information), ERPs elicited by the target polygons
were likely a valid signal to calibrate the classifier. Before each burst, a picture was selected as
the target image and one random polygon from its decomposition (among those with a visual
information above $15\%$) was selected as the target primitive. During this phase, a burst consisted
of $4$ blocks for subjects VPmao, VPmap, and VPjam, and of $10$ blocks for any other subject. After
each burst, a new target image was selected. The subjects were told that, following the display of a
drawing, random polygons would be presented and one or more of them could resemble the drawing or
some salient feature of it. They were not explicitly instructed to seek for these stimuli in an
active way. They were requested to avoid abrupt facial movements, to avoid blinking and to relax the
jaw. Approximately each $15$ minutes a self-paced pause was offered. During the calibration phase,
there was no feedback provided to the participants.

        {\em Reconstruction phase}: We considered all polygons from the decompositions that
contributed more than $3\%$. For each subject, $15$ reconstructions were completed. After choosing
one of the images for reconstruction, it was displayed for $5$ seconds. Then the first burst
proceeded with the most informative polygon of the chosen image as target, together with $5$ 
non-targets drawn from the pool of polygons. Bursts during this phase consisted of $4$ blocks for
subjects VPmai and VPmao, and $10$ for any other subject. During each burst of the reconstruction
phase, the classifier would score the likelihood that each polygon had elicited an ERP, and at the
end of the burst the primitive being most likely the target was determined, as explained in section
\ref{sec:2.04}. This primitive was displayed for $5$ seconds as a feedback alongside the correct
target polygon. Then the original target picture under reconstruction was shown again for $5$
seconds. The next burst proceeded with the next most informative polygon as target and with $5$ new
non-targets from the polygon pool. The correct stimuli from previous bursts were retained after they
had been played out, so that the reconstruction could proceed. After a reconstruction was completed,
a self-paced pause was offered. The instructions to the subjects were the same as during the
calibration phase, except now subjects were told that the reconstruction would proceed until a fair,
schematic rendering of the target image had been reached. A video of a successful reconstruction is
available \cite{videoCherry}.

		\subsection{Data analysis}
			\label{sec:2.04}

      {\em ERP analysis}: EEG signals were lowpass filtered with a Chebyshev filter using a bandpass
up to $40 \> Hz$ and a stopband starting at $49 \> Hz$, and then down-sampled to $100 \> Hz$.
Continuous signals were divided into epochs ranging from $-200 \> ms$ to $1000 \> ms$ relative to
each stimulus onset. Baseline correction was performed on the pre-stimulus interval of $200 \> ms$.
Epochs containing strong eye movements were detected and rejected using the following criterion:
Epochs in which the difference of the maximum and the minimum values in one of the channels F9, Fz,
F10, AF3, and AF4 exceeded $70 \> \mu V$ were rejected. Only those non-target epochs were used in
which the three preceding and the three following symbols were also non-targets in order to avoid
overlap from ERPs of preceding or successive targets. For the grand average the ERP curves were
averaged across all trials and participants. To compare the ERP curves of two classes (target and
non-target) signed $r^2$-values were calculated.

      {\em Classification of what polygons the subject attended to}: We employed binary classifiers
based on spatio-temporal features. We sought for discrimination between epochs related to targets
vs. non-targets. As preprocessing, EEG signals were down-sampled to $100 \> Hz$ by calculating the
average for consecutive data points in non-overlapping stretches of $10 \> ms$ each. Epochs with an
excessive power in a broadband ($5-40 \> Hz$) indicating, e.g. muscular artifacts, were rejected
from the calibration data. The aim of the heuristic is to find five time intervals that have a
stationary (target minus non-target difference) pattern and maximal $r^2$  differences. Occasionally
the intervals determined by the heuristic were adjusted by the experimenter before starting the 
on-line runs (see appendix \ref{app:1.02}). Features were calculated from $55$ channels (all except
for Fp1,2, AF3,4, F9,10, FT7,8; which are the electrodes more exposed to facial movements) by
averaging voltages within each of the five chosen time windows resulting in $55 \times 5=275$
dimensional feature vectors. For classification, a linear discriminant analysis (LDA) with shrinkage
of the covariance matrix \cite{LemmMuller2011} was trained on calibration data. A polygon was
determined by averaging the classifier output for all displayed polygons across the $4$ or $10$
blocks of each burst and then by choosing the primitive that received the largest average output.

      In order to investigate the rate/accuracy tradeoff, the classifier that was used online was
also applied offline to the image reconstruction data. The polygon selection depended then on
averages across the first n blocks, with $n = 1, \dots, 10$. For $n = 10$ the full neural records
are analyzed, so we obtain the same results as when operating the classifier online. For $n < 10$,
some of the data within each trial is discarded, so we are estimating the performance of the
classifier if shorter neural activity records were available. We also computed the Information
Transfer Rate per decision ($ITR_d$) -- i.e. the number of bits involved in each classification --
using:
        \begin{eqnarray}
          ITR &=& log_2(N) + p_c\>log_2(p_c) + \nonumber \\
          && + (1-p_c)\>log_2\left({1-p_c\over N-1}\right), 
          \label{eq:2.04.01}
        \end{eqnarray}
where $N = 6$ represents the number of primitives among which the classifier chooses and pc
represents the empirically measured probability of making the right classification
\cite{MunssingerKubler2014, McFarlandWolpaw2003}. In the offline analysis, by normalizing over the
number n of blocks for $n = 1, \dots, 10$ we get a proxy for the number of bits that the classifier
can extract per block and can thus trade off between the redundant information (offered by the
repeated presentation of the same stimuli) and fast recovery (obtained by less presentations, but
hindered by a lower accuracy). Additionally, we can normalize by the time consumed by a whole burst
to obtain an Information Transfer Rate in bits per seconds (noted $ITR$, without subscript). $ITR$
approximates the amount of information extracted and can be overoptimistic, as noted in
\cite{MunssingerKubler2014}. The $ITR$ is not a realistic performance measure for the BCI context as
it considers redundant coding on the side of the transmitter to achieve error robustness, which is
not a reasonable assumption for the BCI user. In contrast, BCI applications require some mechanism
which allows the user to undo false selections. Since the design of such a mechanism is not
straightforward for our application and requires a full discussion of its own, we still use the
$ITR$ here as a good way to investigate the speed/accuracy trade-off without stressing the absolute
value of the $ITR$.

  \section{Results}
    \label{sec:3}

    \subsection{ERPs}
      \label{sec:3.01}

\begin{figure}[htbp]
				\begin{center}
					\includegraphics[width=0.4\textwidth]{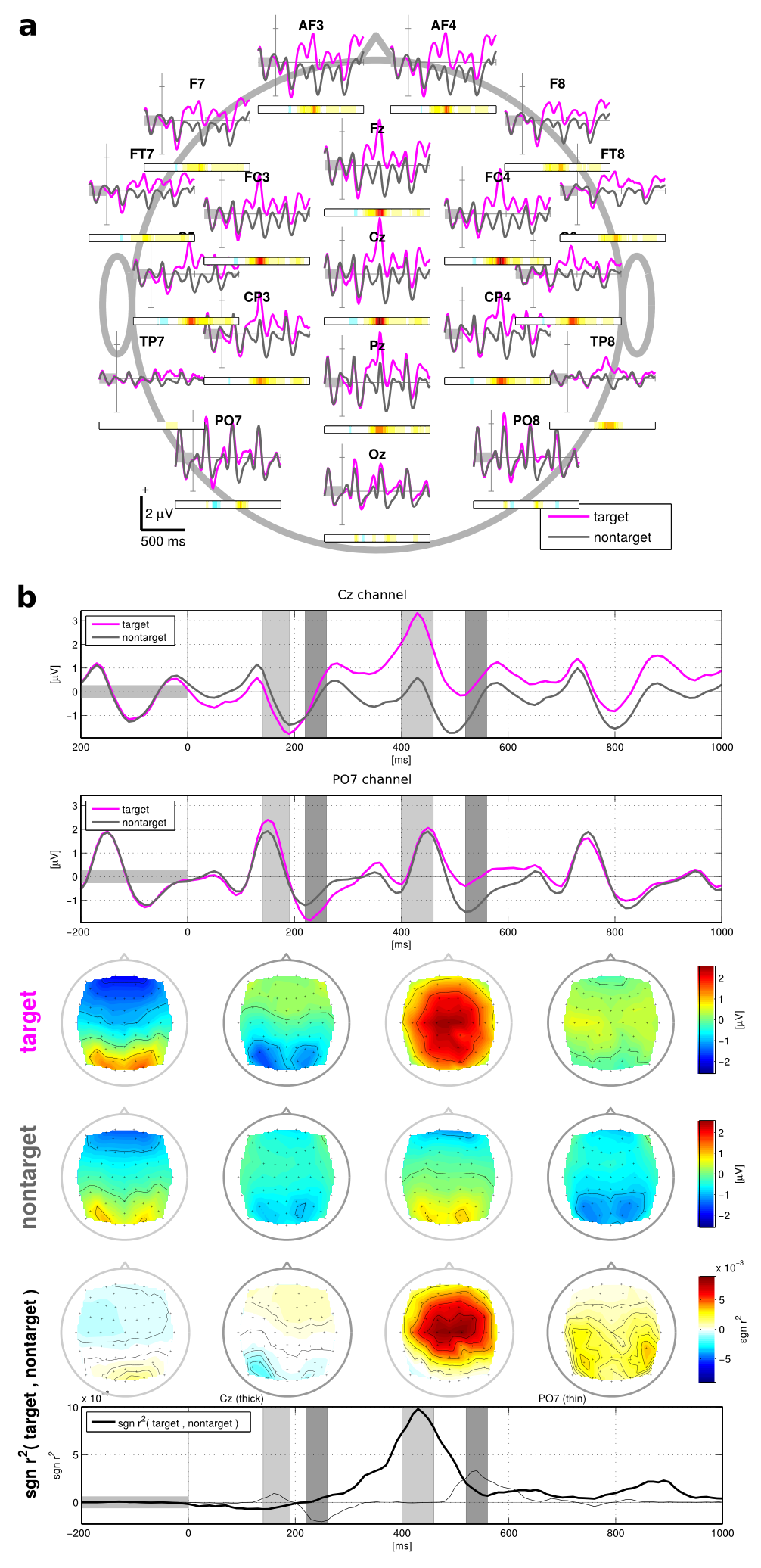}

          \caption{{\bf Event Related Potential analysis}. Signals are averaged over all subjects
and trials. {\bf a} Average activity in different electrodes during target (magenta) and non-target
(gray) stimulus presentation. {\bf b} First and second rows: Averages over subjects of the
potentials at channel Cz (first row) and PO7 (second row) during target and non-target trials.
Shadowed areas delimit important time intervals with notable divergences between signals associated
to target and non-target stimuli. Third and fourth rows: brain activity averaged over each of these
relevant time intervals. Fifth and sixth rows: quantification of divergences between target and 
non-target signals with signed $r^2$-values distributed over the scalp and throughout time as a plot
for the Cz and PO7 channels.}

					\label{fig:3}
				\end{center}
			\end{figure}

      The ERP analysis of the data (figure \ref{fig:3}) presents the grand average of brain activity
over all subjects and trials under target and non-target stimuli. When plotting the scalp map of
this activity (figure \ref{fig:3}{\bf a}) we immediately recognize how the ERPs of interest for our
classification tasks are notable mainly in the frontal and central channels, being the occipital and
temporal areas of null or little interest. This is consistent with the P300 ERP associated to the
oddball paradigm.

      For each subject, time intervals of interest have been selected to provide the classifier with
discriminative data as it was indicated in section \ref{sec:2.04}. In the grand average we can see
how signals usually diverge between target and non-target activity (figure \ref{fig:3}{\bf b}) and
what time intervals are more useful on average; the time interval between $400$ and $500 \> ms$ is
the most discriminative, which is once again consistent with the prominent role that the P300 ERP
should play in the designed BCI.

    \subsection{Classification}
      \label{sec:3.02}

      During the reconstruction phase, the classification implemented after each burst was
considered successful if the classifier had identified the corresponding target polygon from the
current target drawing. This is different to setups where the subjects report whether a stimulus is
target or not \cite{BasaLee2006, ShamloMakeig2009}. Note that in our paradigm a non-target stimulus
might be considered target by a subject and ERPs might be elicited for such stimuli as well, but
these would be scored as wrong classifications if selected by the classifier. This is further
discussed in section \ref{sec:3.04}. On the other hand, different polygons carry different visual
information about the target drawing: making mistakes late in the reconstruction would not be as
dramatic as getting some of the first primitives wrong. We quantified how much of the visual
information was correctly recalled with respect to the maximum possible (note that polygons
contributing less than $3\%$ are never displayed). We refer to this number as {\em weighted
selection accuracy} as opposed to the raw {\em selection accuracy} that measures the percentage of
correct classifications irrespective of their importance.

      \begin{figure}[htbp]
        \begin{center}
          \includegraphics[width=0.5\textwidth]{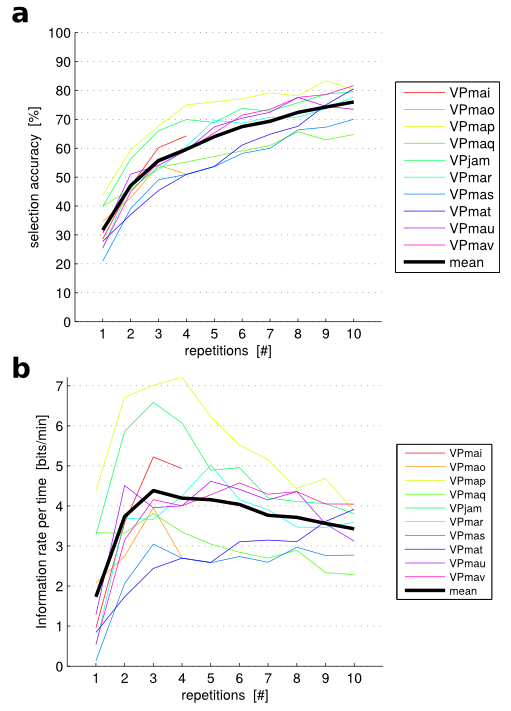}

          \caption{{\bf Offline analysis of performance with alternative settings}. {\bf a} It is
possible to attain acceptable accuracies using less blocks per bursts. The accuracy grows less than
linearly with the number of blocks. This suggests that an optimal operating point might exist at
which a maximal accuracy could be reached faster. {\bf b} $ITR$ per number of blocks reveals a good
operating point at $3$ blocks per burst. This is a very optimistic result, and the approximative
power of $ITR$ is further diminished by the design of the experiment, which does not allow (and does
not require either) for corrections over selected polygons.}

          \label{fig:4}
        \end{center}
      \end{figure}

      The average online selection accuracy across all subjects is $73.4\%$. If we take into account
only those subjects whose classification was based on bursts with $10$ blocks, then the online
selection accuracy raises to $76.4\%$. This must be compared to the chance level for one target
among $6$ stimuli: $16.7\%$. The weighted accuracy is $80.5\%$ ($83.4\%$ for subjects with $10$
blocks).

      In an offline analysis we studied what would be the performance of the BCI if, within each
burst, the classifier would consider less blocks to compute its output, as explained in section
\ref{sec:2.04}. The results are collected in figure \ref{fig:4}. There is a drop in performance for
lower block numbers, as expected, but the average selection accuracy remains relatively high --
above $60\%$ for any choice with more than $4$ blocks per burst (figure \ref{fig:4}{\bf a}).

      In a real free-painting application it would not be necessary to present a template (target)
image. This and the presentation of feedback after each burst affect the time normalization term of
the ITR, whose calculation underlies the plot in figure \ref{fig:4}{\bf b}. For that figure we
assumed $2$ seconds for the display of the selected polygon. We can speculate what would happen
under extremely fast conditions, say $500 \> ms$ for feedback presentation. Then, a peak of $5.4 \>
bits/min$ would be registered also at $3$ blocks per burst (not shown). The opposite, conservative
case with $5000 \> ms$ for feedback (as in our experiment, also not shown) presents its peak at $6$
blocks with an entropy rate of $3.3 \> bits/min$, but the maximum is flat and extends from $3$ to
$8$ blocks per burst.

		\subsection{Performance drop with task difficulty} 
			\label{sec:3.03}

      As indicated in section \ref{sec:2.03.02}, during the reconstruction task a polygon is
selected based on the output of the classifier. This might be the correct polygon -- the one
belonging to the target image reconstruction -- or an incorrect one. Both the correct and the
selected polygons are shown as a feedback to the subject and, irrespective of the outcome, the
correct polygon is held fixed on the background as new randomized stimuli are displayed in the
following bursts. This certainly limits reconstruction freedom, but it allows us to proceed to
tinier details and assess how the accuracy behaves in more complicated scenarios.

      $25.3\%$ of the paintings were completely reconstructed to its full extent without selecting
any wrong polygon. This rises to $28.3\%$ if we consider only subjects whose bursts consisted of
$10$ blocks. (The probability of reconstructing a picture by chance alone is lower than $0.001$ even
for the picture whose reconstruction consists of less polygons). If we would consider the
reconstructions only until the first wrong polygon is selected, a $49.9\%$ ($52.9\%$ with $10$
blocks) of the tasks would have been accomplished across subjects. If we take into account the
visual information conveyed by each polygon and use the weighted accuracy to account for the
percentage of reconstruction complete until the first wrong classification takes place, we find that
$65.3\%$ of the visual information available is correctly retrieved ($68.3\%$ for subjects with $10$
blocks).

      \begin{figure}[htbp]
        \begin{center}
          \includegraphics[width=0.5\textwidth]{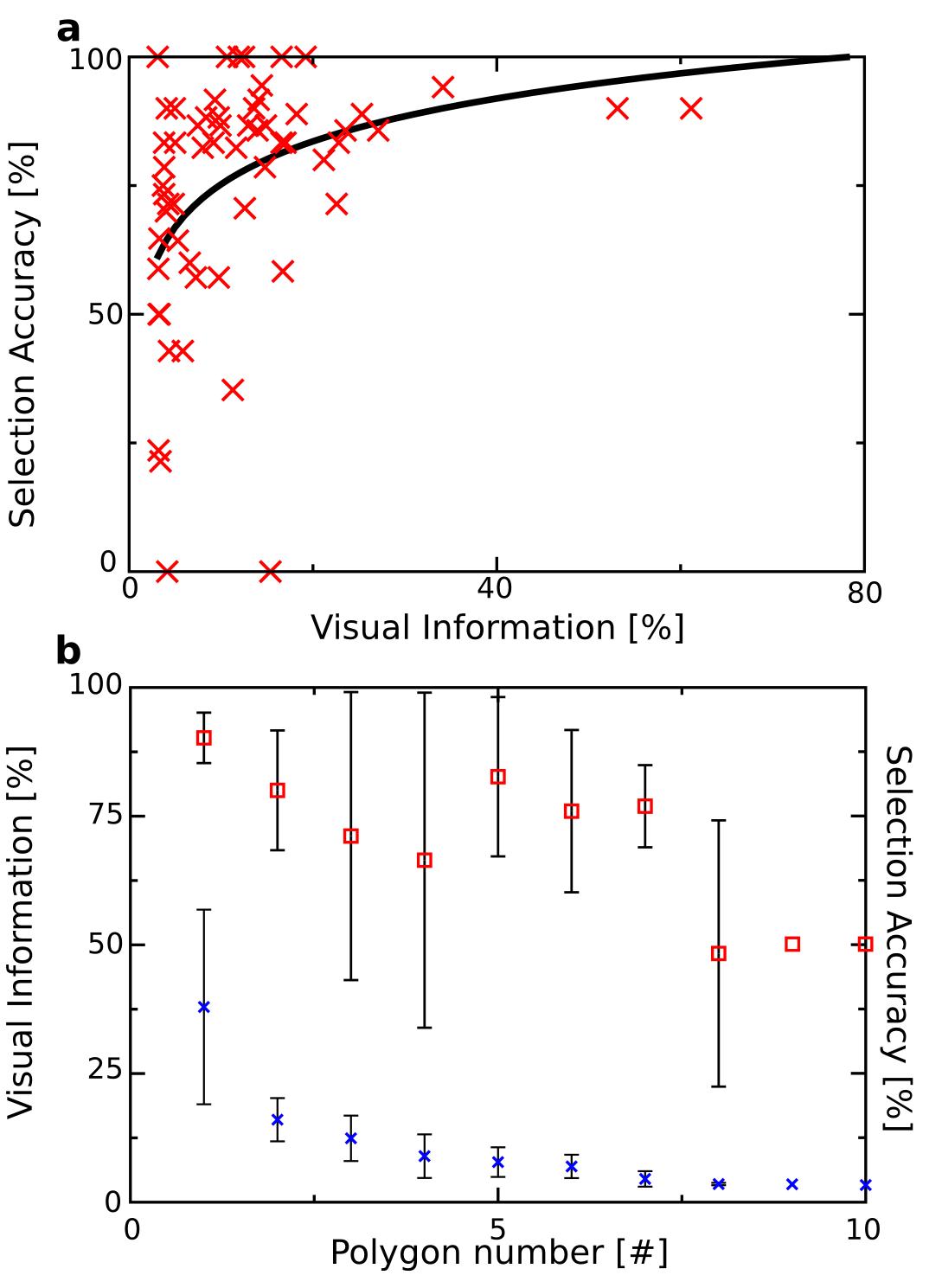}

          \caption{{\bf Performance drop with task difficulty}. {\bf a} Selection accuracy vs.
visual information for individual polygons across subjects. Polygons bearing more information are
usually larger, remind tightly in shape and color to their target, and are consequently easier to
identify. Smaller polygons might reconstruct tiny details or refine existing shapes. These bear less
visual information and represent a greater challenge for the image reconstruction, as indicated by
the drop of performance for less information bearing polygons (thick, black trend line). It is
remarkable, though, that a diversity of polygons with little visual information exists: some of them
are correctly recognized almost always and some others are definitely tough to classify for any
subject. {\bf b} Visual information (blue crosses) and selection accuracy (red squares) vs. rank
that each polygon occupies in the reconstruction of its target. Error bars indicate standard
deviation of the values across polygons. Polygons presented later convey less information and are
consequently more difficult to recognize in average. However, there is a huge variability in
selection accuracy and a marked drop around polygons $4$ and $5$. Both panels were elaborated using
data from experiments with $10$ blocks per burst. There is not any remarkable difference considering
similar plots that include all data.}

          \label{fig:5}
        \end{center}
      \end{figure}

      Because the right polygon was always preserved despite the outcome of the classifier, we could
proceed with each reconstruction until the end and analyze the performance of the BCI as it explores
more complicated scenarios in which target polygons convey very little information (down to $3\%$).
As a result we found the selection accuracy drop for polygons bearing less visual information about
the target (figure \ref{fig:5}). Besides the obvious decay in accuracy for less informative
primitives, we observe an increase in heterogeneity. As we move towards more difficult tasks, a
range of selection accuracies emerge: some less informative polygons are rarely (two of them never)
recognized as part of the target image while others are still correctly selected to a great extent.
Polygons bearing more visual information tend to be accurately selected most of the time. Only $6$
polygons have always been correctly selected, all of them had a visual information below $20\%$ and
one of them is at the edge of the $3\%$ threshold. These results were obtained for the eight
subjects with $10$ blocks per burst during the reconstruction task. The outcome considering all
subjects is broadly the same.

      In accordance with the experimental design, less informative polygons appear later in the
reconstruction task, as shown in figure \ref{fig:5}{\bf b}. The selection accuracy almost always
decays as a reconstruction proceeds. Remarkably, the selection accuracy for the first polygon is
$87.3\%$ ($90.8\%$ for subjects with $10$ blocks per burst). In figure \ref{fig:5}{\bf b} we also
see an unexpected drop (followed by a rise) of the selection accuracy for intermediate polygons. The
reason for this is discussed in appendix \ref{app:2} and relies on the particularly difficult task
that polygons $3$ or $4$ of some reconstructions posed to the subjects.

    \subsection{Ambiguous polygons}
      \label{sec:3.04}

      In choosing our pictures for reconstruction, we want them to be iconic with easily
recognizable parts and with as little overlap as possible. However, it cannot be avoided that some
drawings (or parts of them) resemble some other one. This leads to an undesired effect during the
experimental sessions: some polygons did not belong to the reconstruction of the target drawing, but
they bore some resemblance to it and were often classified as target. Because these polygons are
strictly non-targets, they were excluded from the selection accuracy results reported in section
\ref{sec:3.02}. This does not imply a malfunction of the classifier because such polygons could have
tricked the subjects as well. These pieces introduce an interesting ambiguity as they could
contribute to several reconstructions. We refer to them as {\em ambiguous polygons}.

      \begin{figure}[htbp]
        \begin{center}
          \includegraphics[width=0.5\textwidth]{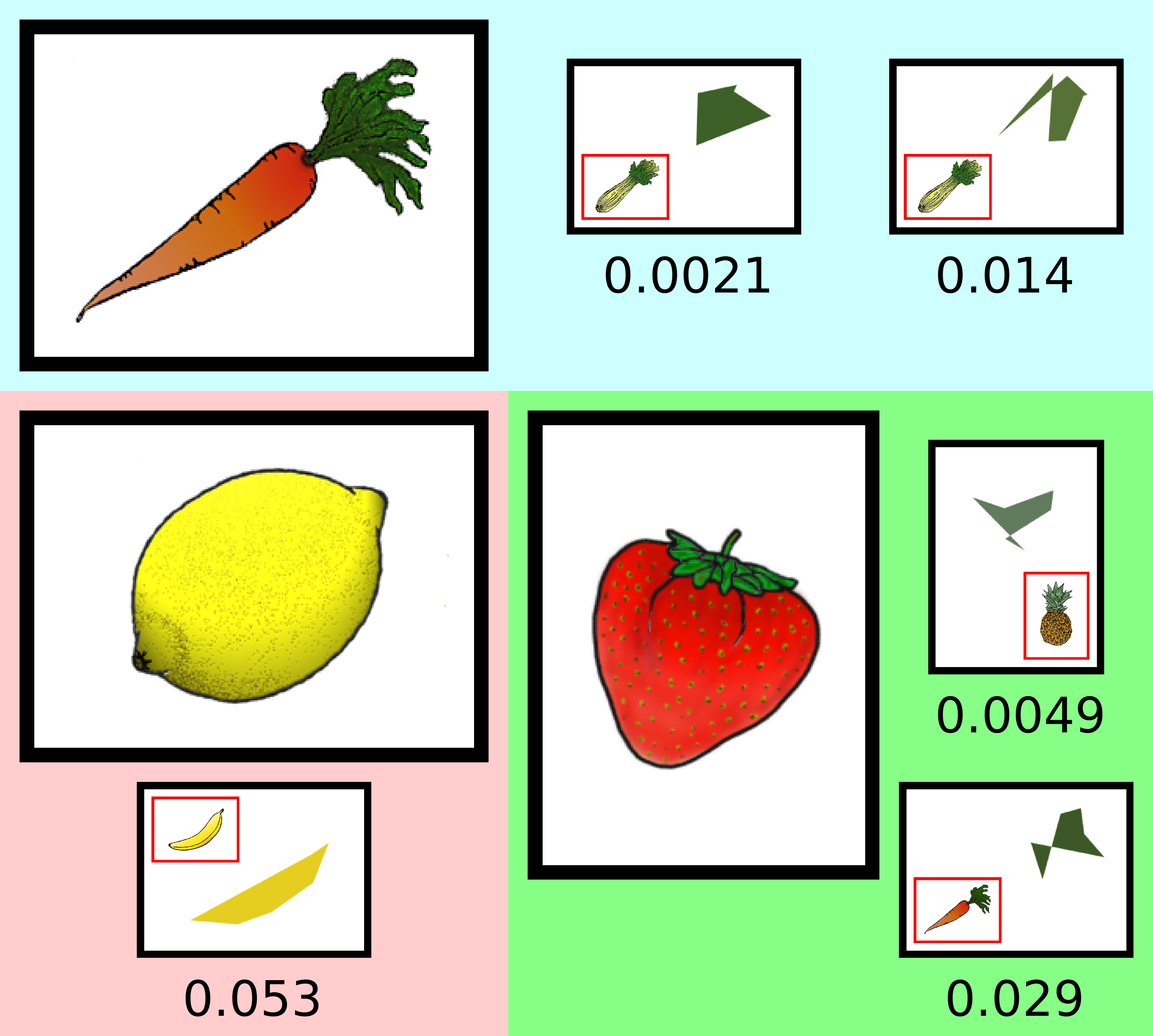}

          \caption{{\bf Ambiguous polygons}. Some primitives could contribute to the reconstruction
of different images. This could interfere with the reconstruction since these polygons would not be
scored as correct classifications. The ambiguity of these pieces -- that stems from the visual
overlap between different targets -- could be exploited to speed up the BCI image reconstruction. We
shown the carrot, lemon, and strawberry together with some polygons that do not belong to their
decompositions but that are overly selected by the classifier. For each polygon, within a red frame,
it is shown the original drawing to which they belong. This figure was elaborated using data from
experiments with $10$ blocks per burst. The scenario is similar if we use all data for this
analysis.}

          \label{fig:6}
        \end{center}
      \end{figure}

      As explained in section \ref{sec:2.03.01}, along with each target polygon we chose $5$ 
non-target polygons from one common pool. One same polygon might have shown up several times as a
non-target during one reconstruction, or for the reconstruction of the same drawing by different
subjects. If in such cases the classifier repeatedly selected the non-target, we might have strong
evidence that the subject perceives that polygon as contributing to the target drawing. Based on
this, we computed $p$-values to ascertain what polygons had been more probably not selected by
chance alone, but presumably because an honest interference existed between those non-targets and
the target drawing. We show the top $5$ such pieces in figure \ref{fig:6} along with their
$p$-values and the target images for which they were significantly over-selected. We see that these
could perfectly contribute to sketch the target (like the polygon from the banana in the case of the
lemon reconstruction) or to the refinement of a smaller detail (like the green polygons seemingly
selected to complete the leaves of the carrot and the strawberry). In these paradigmatic cases the
wrongly classified non-targets would show up roughly in the same area as the actual target polygons.

	\section{Discussion}
		\label{sec:4}

    In this paper we show how Rapid Serial Visual Presentation using bursts of polygons together
with the oddball paradigm can be used for BCI image reconstruction. Our purpose was merely to attain
a proof of concept. For that end, different design choices have been made and tested during the
experimental sessions. A systematic search of the best working settings was never intended and is
left for future work. Notwithstanding this, the results reported in section \ref{sec:3} invite us to
be optimistic about the paradigm and demand that further, more rigorous experiments be performed. In
this section we comment on our results and compare them to previous approaches to BCI-painting. We
also propose future lines of research focusing on the development of a free-painting BCI based on
the current paradigm.

    \subsection{Classification accuracy, performance drop, and comparison with previous research}
      \label{sec:4.01}

      In section \ref{sec:3.02} we reported a classification accuracy of $74.4\%$ ($76.4\%$ for
subjects with $10$ blocks). This rises to $80.5\%$ ($83.4\%$) if we acknowledge that some correct
classifications are more important than others, and use the percentage of visual information
retrieved to weight our results. This means that we can reproduce up to $80\%$ of the visual
information that the polygon decompositions capture of the original images.

      These numbers would gain relevance if compared with those of other BCI-painting paradigms. We
attempt this now, but a series of limitations exist due to the differences in BCI designs. For
example, the guided evolution in \cite{BasaLee2006} towards images that produce `positive' feelings
cannot be properly quantified. The authors report a $61\%$ accuracy, but this might be a biased
result. After three generations of the genetic algorithm (which takes around $36$ minutes), subjects
in [13] were asked to recall their actual choices -- those picture that elicited more `positive'
feelings according to their conscious experience. Unfortunately, this approach is biased: the image
selected by the classifier is used to generate variations that are then persistently shown to the
subject, while the un-selected pictures are lost. One further limitation to compare our results to
this work was pointed out at the introduction: in \cite{BasaLee2006}, there are no correlates with
actual visual structures. While we quantify the overlap of our reconstructions with the original
drawing (though naively, through the visual information), pictures in \cite{BasaLee2006} evolve
towards positive-looking images according to physiological feedback. Guiding the evolution towards
visual details that we could quantify is rather difficult.

      The impressive classification accuracy obtained in \cite{ShamloMakeig2009} clearly outperform
ours. This work shall be a good reference for future BCI-painting paradigms, especially if it is
possible to attain their high presentation rate ($8$ images per second during each burst) while
keeping up in selection accuracy when the design is tested in actual image evolution tasks.
Comparison with other aspects of our work is more difficult and we have to wait until this image
evolution paradigm is put to an online test. Then, we will be able to study a sense of {\em
progression} towards a final target and we could quantify properly how well detailed visual
arrangements get reconstructed. If the accuracy results reported in \cite{ShamloMakeig2009} persist,
this is a truly promising scheme.

      Finally, the {\em P300-Brain Painting BCI} \cite{KublerHosle2008, HalderKubler2009,
MunssingerKubler2014} performs very well in selection accuracies and it also provides two measures
of $ITR$, reporting values that beat our BCI design. This is a very promising paradigm that mimics
the equally successful matrix spellers. However, a few points should be considered. In usual
spellers, a sentence is provided for a subject to copy. In \cite{MunssingerKubler2014}, a picture is
produced if we follow a series of instructions precisely: these indicate what shape to place next on
the canvas, what color and size it shall be, and where it must be placed -- all these instructions
are stored as symbols in the matrix interface. These are the symbols that are accurately retrieved
from the matrix for the results in \cite{MunssingerKubler2014} to make sense. Hence this is an {\em
instruction-copying} task rather than a {\em copy-painting} one. If, otherwise, a subject were
provided an image and were requested to copy-paint it, there would be several different sets of
instructions leading to the same outcome, and such variations are more complicated to account for in
terms of symbol accuracy. A similar problem restricts our BCI so that we have to dismiss ambiguous
polygons, as explained above. While this implies some limitations for our paradigm, the problem is
lightly deeper in \cite{MunssingerKubler2014}: copying a set of symbols (instructions) does not
necessarily relate to visual cues on the canvas and, accordingly, makes quantification of the result
more difficult. Specifically, it is not straightforward to find a meaningful measure equivalent to
our visual information. While our selection accuracy and, specifically, our weighted accuracy report
directly about the objective overlap that we attain with an actual target drawing, the results in
\cite{MunssingerKubler2014} are more difficult to quantify in visual space.

      We expect that future research on BCI-painting will make possible a more systematic comparison
between different paradigms and visual outcomes. There is, though, a relevant caveat of our own BCI
that deserves a closer analysis. In section \ref{sec:3.03} we report a performance drop as the
difficulty of the task increases. Because of this, only $25\%$ of the reconstructions could proceed
as intended. This should be improved in future implementations of the paradigm, and hence a series
of alternatives are available:
        \begin{itemize}

          \item The rather conservative experimental settings (SOAs, some display options for the
polygons, number of blocks per burs, etc) make possible an improvement to achieve faster $ITR$ and
equally better accuracies. Gaining a few seconds per burst could allow us to introduce some error
correcting mechanism, as discussed below, while keeping a high $ITR$.

          \item The classification results in \cite{ShamloMakeig2009} might be traced back to an
easy cognitive task: $94\%$ of the bursts containing a target are consciously detected by the
experimental subject, perhaps because the classification involves a clear-cut task. In our case (due
to the presence of ambiguous situations) conscious wrong decisions might be prominent. This strongly
suggests that to improve our accuracy we should pay more attention to the input stimuli. We could
either exploit the role of ambiguous polygons or attempt to ban them altogether by building
orthogonal primitives that produce a maximum diversity of designs with a minimum number of patterns.
An interesting alternative could be to separate the generation of new shapes and their filling with
color. This is an open and interesting problem at the frontier between BCI and natural image
decomposition.

        \end{itemize}

      Admittedly a $25\%$ complete reconstructions falls short, but we suggest that this is not the
best indicator of the performance of the BCI. It is desirable to complete the reconstruction of an
image but this is a very stringent condition: finer grained primitives become more difficult to
classify, while they might not be as important as the overarching primitives. Tinier details might
also be open to more subjective appreciations by the BCI users. Furthermore, towards the end of a
reconstruction the quality of the current primitives might degenerate since the genetic algorithm
favors convergence of broad details first. We believe that weighting the classification accuracy by
the amount of visual information that each classification contributes is a better indicator of the
performance of our BCI paradigm. Then, as reported in section \ref{sec:3.02}, up to $65.3\%$
($68.3\%$ for subjects with $10$ blocks per burst) of visual information can be retrieved before the
first classification mistake.

    \subsection{Moving towards a free-painting BCI}
      \label{sec:4.02}

      If we wanted to move towards an RSVP-based free-painting BCI machine, we could exploit the
existence of ambiguous primitives and we should consider seriously the necessity of error correcting
mechanisms. Other improvements should direct the evolution in an active way, avoiding full reliance
on the randomness of primitives as discussed below, e.g., for the $2$-D location of new polygons. We
speculate about the future of free-painting BCIs based on our paradigm in the following.

      We consider first the possibility of including error correcting mechanisms to improve the
classification accuracy and $ITR$. In computing the speed/accuracy trade-off by the $ITR$ formula,
we neglected the problem of explicit correction of false BCI selections. These corrections that
would have to be performed explicitly by the user are more time consuming than being accounted for
by that equation (which assumes optimal error robust coding). Accordingly, a realistic trade-off
would result in a higher number of blocks than the three suggested by figure \ref{fig:4}{\bf b},
reflecting the delay introduced by error correction. Other RSVP tasks have mechanisms to correct for
wrong outputs. In RSVP spellers, a symbol can be explicitly incorporated among the letters to
represent the backspace \cite{FarwellDonchin1988}. A similar approach could be taken for the present
BCI application but it would likely interfere with the display of polygons. An appealing alternative
are error potentials, a large scale signal elicited by unexpected feedback after a classification
task, which can be used to automatically cancel incorrect selections \cite{SchmidtTreder2011}.

      One further alternative is inspired by \cite{LorenzVidaurre2014} where ERP-based BCIs are
explored. The selection accuracy is well characterized and a confirmation step (based on a similar
RSVP task) is analyzed rendering a $96.26\%$ success with no false positives. Incorporating one such
confirmation step would raise our selection accuracy to $82.1\%$. For this we computed the
probability (across all subjects) that the second highest ranked polygon was the correct one given
that the first choice was a wrong classification -- this spares us a second burst but has a lower
accuracy. If we would incorporate two confirmation steps (and, consequently, use the probability
that the correct polygon was the third highest ranked one, provided the two first were not), the
selection accuracy would rise to $87.8\%$. Alternatively, we could proceed with a second burst
discarding the first (wrongly selected) polygon and introduce a new non-target one. This way, the
accuracy would reach a $90.2\%$ for just one confirmation step. To include these details in an
estimation of the $ITR$ would be very speculative: we should take into account not only the
increased time lapse of the error correction task, but also potential undesired effects due to the
disruption of the main painting task.

      Before using solutions that need more steps, as the error correcting mechanisms discussed, we
note that some available, relevant information is dismissed in the current design. A 
winner-takes-all decision is forced at the end of each burst while it is possible that several
polygons get large scores from the classifier, especially if ambiguous primitives are present. On
the other hand, easier classification at the beginning of a reconstruction might have a clear winner
early in the burst. We could quantify the uncertainty of a decision (e.g. through an entropy-based
classifier) and exploit this information, which is already captured by our BCI. The length of a
burst could be dynamically tuned until the confidence of a decision would rise above a threshold. A
sustained uncertainty could be taken as `select nothing', an interesting option that free-painting
applications should allow.\\

      Regarding ambiguous polygons, these are pieces that might contribute to the reconstruction of
several different images as reported in section \ref{sec:3.02}. Usually, this is because the
original drawings themselves share some common traits, as in the case of the green leaves located
around the same position for the carrot, the strawberry, etc (figure \ref{fig:5}). They do not
represent a generalized situation: it would just affect a few polygons, and scoring these cases as
correct classifications would not change our selection accuracy significantly.

      For our results, these ambiguous polygons had a negative effect because they do not count as
correct classifications (and thus lower our selection accuracy); but in a wider scope they might be
extremely useful. While we imposed that the image decomposition be unique for each drawing, we can
conceive of intermediate decompositions of multiple images with shared primitives. Then, a
reconstruction should proceed from ambiguous, generalized descriptions towards more particular ones.
This would establish a hierarchy that would cluster together pictures that are closer to each other
in visual terms. By exploiting this feature we could discard non-targets quickly if they are very
distant from our target but, as a consequence, we will progress towards more difficult
classification tasks. To work out this situation we must research what is the finer detail that our
BCI paradigm is able to resolve.

      Linked to this, we could seek the use of maximally discriminating primitives at each
reconstruction stage. Note the few constraints that we imposed upon the bursting polygons: many
might be displayed within the same burst that convey redundant information, thus diminishing the
exploratory capabilities of RSVP. Also, shape and color are tightly linked together in the current
design. A two-stage decision that separates these arguably orthogonal features might be of great
help towards a free-painting BCI. How we should design our primitives to maximally exploit the
current paradigm is one of the open research lines proposed for the future.\\

      Finally, there is an important aspect that is left to random chance in the current paradigm
that should be corrected, if we really wanted our BCI users to explore freely the space of possible
drawings. Right now, correct polygons are placed where they belong because their location is part of
the specifications of each polygon decomposition. If we did not know the correct location of a
primitive, it is extremely unlikely that it would be placed at the right spot by chance alone.

      We envision the next design: Instead of the whole screen, consider a bursting area, a
rectangle of restricted size laid upon the canvas within which all bursting polygons fit.
Accordingly, the drawing will only suffer modifications in the framed area. The position and size of
the bursting area could be controlled by a joystick, but this would not be appropriate for impaired
users. Instead, the position of the frame could be modified through an eye tracking device or by
sensory motor rhythms (as proposed in \cite{MunssingerKubler2014} for the {\em P300-Brain Painting
BCI}). This last option would be the preferred choice for patients who cannot focus their sight
(e.g. those suffering completely locked-in syndrome), but in this last case the drawing beneath the
bursting area should be displaced while the bursting area remains fixed under the focus of the BCI
user. Note that bursting is halted while the bursting area is manipulated.

      To control the size of the bursting area when a joystick is not a viable option, we propose
using the edges of the canvas as part of the interface: bringing the bursting area to the left-most
edge of the canvas and insisting that it moves further than allowed would enlarge the horizontal
dimension of the frame, while moving it all the way to the right and insisting that it goes further
(e.g. focusing the sight outside the canvas region) would diminish the horizontal dimension. The
same scheme would work for increasing/decreasing the vertical dimension using the top and bottom
edges respectively, and the top-left and bottom-right corners would operate on both dimensions at
the same time. We propose also that the top-right and bottom-left corners could be used to zoom in
and out the canvas -- note that this is strictly different from enlarging/shrinking the bursting
area, since it allows for control over ever tinier details while the bursting polygons appear large
to the BCI user.

      Locating polygons in depth seems to be the remaining challenge. If necessary, we could dismiss
the simultaneous modification of horizontal and vertical dimension and use the top-left and 
bottom-right corners to increase/decrease the layer where new primitives are bursting.

	\section*{Acknowledgments}

    This work was supported in part by grants of the BMBF: 01GQ0850 and 16SV5839. The research
leading to this results has received funding from the European Union Seventh Framework Programme
(FP7/2007-2013) under grant agreement 611570. We thank two anonymous reviewers for very constructive
comments and Edward Zhong for his valuable revision and corrections of a late draft of the paper.
Seoane acknowledges the support of the Fundaci\'on Pedro Barri\'e de la Maza and useful discussion
with members of the Complex Systems Lab, especially Sergi Valverde.

  \appendix

  \section{Discussion of important design choices}
    \label{app:1}

    \subsection{Preprocessing of target images}
      \label{app:1.01}

      Both writing and painting are emergent processes, although at very different levels. While in
the former minimal components are clearly identified (written characters, letters) the latter is a
truly emerging outcome of the very strongly, non-linearly interacting pieces that compose an image.
It may be impossible to come down to some basic components of a drawing. If we intend to produce a
picture from scratch, our choice of building blocks (say our {\em alphabet} for drawing) conditions
the complexity that can be generated, the difficulty required to render each picture, and the speed
to which we can produce it. We need to find adequate primitives that can compose a range of images
quickly by combining the minimal units. Our pieces should be simple and schematic, and as pivotal to
our targets as letters are to writing. We think that this is an open problem. For this study we
adopted a provisional solution that we describe in the following.

      As targets for reconstruction we sought iconic images from the Snodgrass and Vanderwart's
object database \cite{RossionPourtois2001}. The chosen pictures (figure \ref{fig:1}) are drawings of
fruits and vegetables with basic shapes and colors; all of them laid on a white background, so that
the reconstruction focuses in clear motives and not in peripheral details. Note anyway that the
subject's attention is free to wander over the screen. We discuss subject focus again in section
\ref{app:1.02} in comparison with previous RSVP applications.

      We need schematic, yet faithful, representations of the selected pictures. That was a
preprocessing step completed weeks before the experiments. To extract useful primitives from our
pool of drawings we found the perfect tool in recent applications of genetic algorithms (GAs) for
image decomposition \cite{genAlgorithm}. Such algorithms proceed through mutation and artificial
selection from an arbitrary population of polygons to a set of polygons carefully arranged as to
mimic a desired image.

      For a GA, a fitness function needs to be introduced. Take a set of random polygons laid upon a
white background, and some on each other to create a sense of depth. These compose an arbitrary
image. Given an RGB color scheme, we use as fitness function the pixel-by-pixel Euclidean distance
between the original picture and the image rendered by the collection of polygons. The better the
fitness, the closer the collection of polygons resembles the original image.

      As a seed for the algorithm we use an arbitrary collection of polygons $p_{i=0} = \{p_{0,j}; j
= 1, \dots, n_i\}$, where $i$ labels iterations and $j$ labels the $n_i$ polygons composing the
collection. Note that $n_i$ might change from one iteration to another. At each iteration the
fitness of the current collection $P_i$ is evaluated. Some mutations are applied to generate an
alternative polygon composition $P'_i$. The collection with better fitness is retained: $max_F \{
F(P_i), F(P'_i) \} \rightarrow P_{i+1}$, where $F(x)$ stands for the fitness function. The algorithm
continues until a satisfactory convergence towards the original image has been reached or until the
fitness does not improve for several iterations. The collection $P_{i_{end}}$ when the algorithm
halts is referred to as the {\em polygon decomposition} of the original image.

      The possible mutations applied at each iteration are: removal of a polygon or insertion of a
new random one; swapping two polygons (recall that some polygons overlay some others); random
addition, deletion, or modification of polygon's vertices; change of a polygon color. Details on the
probability of each operation (provided along the code \cite{codeGA}) are not relevant as long as a
satisfactory approximation of the original images is achieved (which was done, as appreciated in
figure \ref{fig:1}).

      The algorithm allows options regarding the kind of polygons that could be used. This turned
out to be very important. We wished to discard very complex pieces; therefore only polygons with $3$
to $7$ vertices were allowed. We did not use partially transparent polygons -- as implemented by the
$\alpha$ parameter in the RGB color scheme. This was a valid option for the original GA
\cite{genAlgorithm}, but it introduced important non-linearities in the interactions between
polygons. As an instance, if transparency were allowed it might happen that some desired color shows
up only after two polygons have been stacked in the right position. If only one of the polygons were
present, we would not get quite the exact shade. But our paradigm only allows one polygon at a time,
so we would risk rejecting the right polygon because it does not show the final result yet. By using
only opaque polygons we partially solved this problem. \\

%%%%%%%%%%%%%%%%%%%%%%%%%%%%%%%%%%%%%%%%%%%%%%%%%%%%%%%%%%%%%%%%%%%%%%%

      Once an image has been decomposed in its primitives, the fitness function also offers a
measure of the importance of each piece in the reconstruction. For each polygon $p_j \in
P_{i_{end}}$ we compute the fitness function of the whole arrange of polygons, and the fitness of
the same arrange when $p_i$ is removed. We defined the {\em visual information} carried by the
polygon as the normalized drop in fitness. There is no intention of connecting this visual
information to actual theoretical information measures.

      Using this visual information we ranked the polygons for each of our original images and
retained only those contributing more than a $3\%$ (or $15\%$, see section \ref{sec:2.03.02}). This
choice renders fine enough reconstructions (see figure \ref{fig:1}). Adding more polygons would only
help with very tiny details and would make the experiments unnecessarily tedious.\\

      The oddball paradigm requires that target stimuli are presented intermixed with neutral
stimuli. When an original image from our pool was chosen as the object for reconstruction, all the
polygons in its polygon decomposition became target stimuli. As for neutral stimuli, we used a pool
containing all the polygons belonging to the decomposition of all other drawings.

      This decision made the reconstruction task a fair one. All the polygons in our pool of neutral
stimuli have been generated following the same procedure, only they belong to the polygon
decomposition of different images. Differences between the two classes would ideally arise from
their belonging or not to the target decomposition. If we would use, e.g., completely random
polygons as neutral stimuli these could take any aspect, obviously including shapes that would
hardly contribute to the generation of natural images. Such instances could be readily identified
when opposed to polygons that account for details of some natural image. The reconstruction task
would be artificially simplified by a preselection that greatly reduced the uncertainty about the
target primitives. The task would effectively become a {\em recognition of the less random- looking
polygon}.

    \subsection{Experimental setup}
      \label{app:1.02}

      The natural guidelines for our BCI design are the experiments with RSVP spellers by Acqualagna
and Blankertz \cite{AcqualagnaBlankertz2010, AcqualagnaBlankertz2013}. These give us a vague idea of
settings under which a BCI for image reconstruction could function. We chose rather conservative
specifications, given the novelty of the procedure. For example, during RSVP a burst SOA of $330 \>
ms$ between consecutive polygons was used. This lapse includes $100 \> ms$ during which a void
stimulus was intercalated. These settings can be compared to the SOAs of $116$ to $83 \> ms$ from
\cite{AcqualagnaBlankertz2010, AcqualagnaBlankertz2013}, where letters succeeded each other without
inserting any voids. These differences (together with the good results reported here and those from
RSVP spellers) suggest that there is room for improvement should we seek more ambitious experimental
settings.

      The original motivation to introduce RSVP to BCI spellers was that letters could be presented
always at the focal point. Stimuli falling right in the visual focus of the subject show enhanced
ERPs which are detected more easily. This is not an asset in our case: we must allow the subjects to
focus on different areas of their visual field when peripheral details of the target images are
being reconstructed. We did not impose any conditions on the focus of the subjects. Looking at the
long term goals of this research, the interest of RSVP will be on its exploratory potential because
of the random, combinatorial nature of the bursting primitives and its interplay with the subject's
(sub)conscious driving of the image reconstruction process. We are convinced that these ideas are
worth exploring.

      As indicated in section \ref{sec:2.04}, occasionally, the intervals determined by the
heuristic were adjusted by the experimenter before starting the on-line runs. These adjustments
where generic for all subjects and intended to guide the classifier towards informative time
intervals. In all cases, the linear classifier converged towards five discriminative temporal
intervals (ideally towards the most discriminative ones), partly erasing the manual setup. Table
\ref{tab:1} collects the eventual time intervals chosen by the classifier for each subject.

      \begin{table*}[]
        \centering

          \caption{Discriminative time intervals found by the linear classifier for each subject.}
          \label{tab:1}

          \begin{tabular}{| l | c | c | c | c | c | }
            \hline
            BCI subject: & $T_1$ [$ms$] & $T_2$ [$ms$] & $T_3$ [$ms$] & $T_4$ [$ms$] & $T_5$ [$ms$] \\
            \hline
            VPjam & $70 - 120$ & $120 - 150$ & $240 - 350$ & $345 - 480$ & $475 - 530$ \\ 
            \hline
            VPmai & $110 - 160$ & $360 - 390$ & $385 - 500$ & $495 - 590$ & $585 - 810$ \\ 
            \hline
            VPmao & $210 - 290$ & $340 - 440$ & $435 - 560$ & $555 - 570$ & $640 - 720$ \\ 
            \hline
            VPmap & $100 - 130$ & $160 - 200$ & $195 - 280$ & $275 - 470$ & $465 - 530$ \\ 
            \hline
            VPmaq & $220 - 400$ & $395 - 440$ & $435 - 650$ & $645 - 680$ & $680 - 760$ \\ 
            \hline
            VPmar & $130 - 170$ & $210 - 390$ & $385 - 480$ & $475 - 590$ & $585 - 690$ \\ 
            \hline
            VPmas & $240 - 350$ & $450 - 500$ & $550 - 620$ & $615 - 680$ & $675 - 770$ \\ 
            \hline
            VPmat & $220 - 250$ & $245 - 280$ & $275 - 370$ & $365 - 510$ & $505 - 790$ \\ 
            \hline
            VPmau & $210 - 250$ & $290 - 320$ & $315 - 450$ & $445 - 560$ & $600 - 660$ \\ 
            \hline
            VPmav & $130 - 180$ & $280 - 330$ & $325 - 470$ & $465 - 590$ & $585 - 610$ \\ 
            \hline
          \end{tabular}
      \end{table*}

      For the present proof of concept, we focused on achieving a classification accuracy good
enough, for which we took into account as much spatio-temporal information as possible. Underlying
this choice are the patterns inferred by the classifier, which involve the activity of most channels
available during a series of time intervals (see Table \ref{tab:1}). However, the current design
relies strongly on the P300 oddball paradigm whose macroscopic imprint is a frontal/central positive
deflection of the scalp potential. More classic accounts of this signal rely on measurements on
central electrodes alone, prominently on the Cz channel.

\begin{figure}[htbp]
        \begin{center}
          \includegraphics[width=0.5\textwidth]{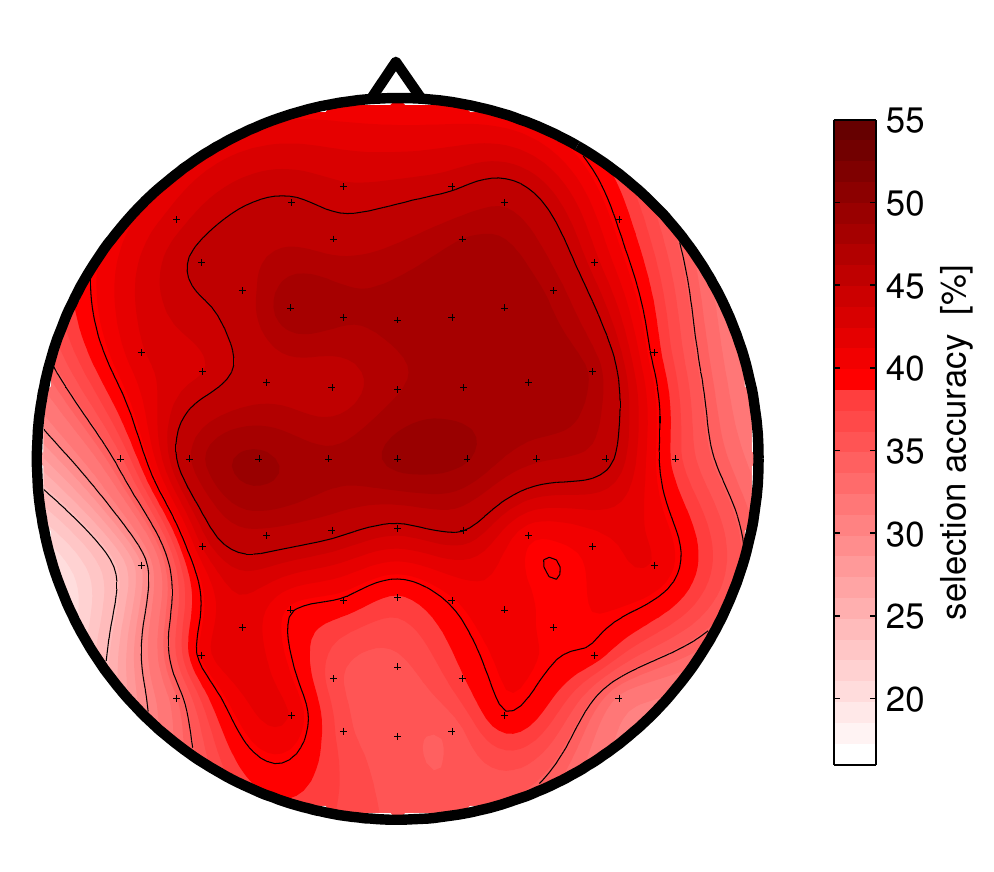}

          \caption{{\bf Selection accuracy scalp plot for a classifier trained on individual
channels alone}. This image reveals that the most informative channels occupy fronto-central
positions (remarkably around the Cz channel), which is in accordance with a prominent role of the
P300 ERP in the task under analysis. However, a notable drop in comparison with the accuracy of
extended spatio-temporal filters indicates either that signals other than P300 are valuable, or that
that signal is enhanced through implicit spatio-temporal filtering when data from many channels is
available. We cannot decide between both possibilities currently.}

          \label{fig:7}
        \end{center}
      \end{figure}

      It is fair to ask whether these central channels alone have discriminative power to accomplish
the image-reconstruction task of our BCI. In figure \ref{fig:7} we plot the average selection
accuracy per channel if that channel alone were considered by the classifier. We observe that the
central and fronto-central channels are the most predictive ones, strongly supporting that the P300
fronto-central potential underlies the analyzed ERP.

      We also observe a notable drop in classification accuracy compared to the quality obtained
with compound spatio-temporal filters (below $50\%$ for all individual channels compared to $70\%$
and higher accuracies reported in the body of the paper). This might suggest a relevant role for
other sources than the fronto-central channels. Notwithstanding, this does not imply that the
enhanced accuracy of extended spatio-temporal filters necessarily stems from signals other than the
P300 deflection. Due to the availability of data from multiple channels, the signal of interest may
be enhanced by implicit spatial filtering. With the available data, we cannot distinguish between
these possibilities.

  \section{Average performance drop for intermediate polygons}
    \label{app:2}

    In section \ref{sec:3.03} it was reported the performance drop as the difficulty of the
reconstruction task increased. Two approaches were taken: i) The selection accuracy was plotted
against the visual information carried by each polygon and ii) the selection accuracy was plotted as
a function of the rank that each polygon occupied in the reconstruction. The first method asserts
that the selection accuracy drops in average for polygons that contribute less to the reconstruction
of the image (figure \ref{fig:5}{\bf a}). Highly informative polygons are usually correctly
classified, but the least informative polygons display a great variation. Some of them are often
well classified and some of them are not, with a range of selection accuracies in-between. This
indicates that the reconstruction task does not always become more difficult as we move towards less
informative polygons.

    Figure \ref{fig:5}{\bf b} reports the average selection accuracy across all reconstructions
against the rank that a given polygon occupies in the reconstruction. The reconstruction task
proceeds from the most informative polygons to the least informative ones, given the target picture.
The latter contain less information about the original image, and hence it should be more difficult
to classify them correctly. Unexpectedly, the decay in performance is not a monotone function:
polygons $5$, $6$, and $7$ in the reconstruction are selected with notably more accuracy than
polygons $3$ and $4$. This indicates that there are some less informative polygons that are
correctly selected more often than the average of other, more informative polygons.

    \begin{figure}[htbp]
      \begin{center}
        \includegraphics[width=0.5\textwidth]{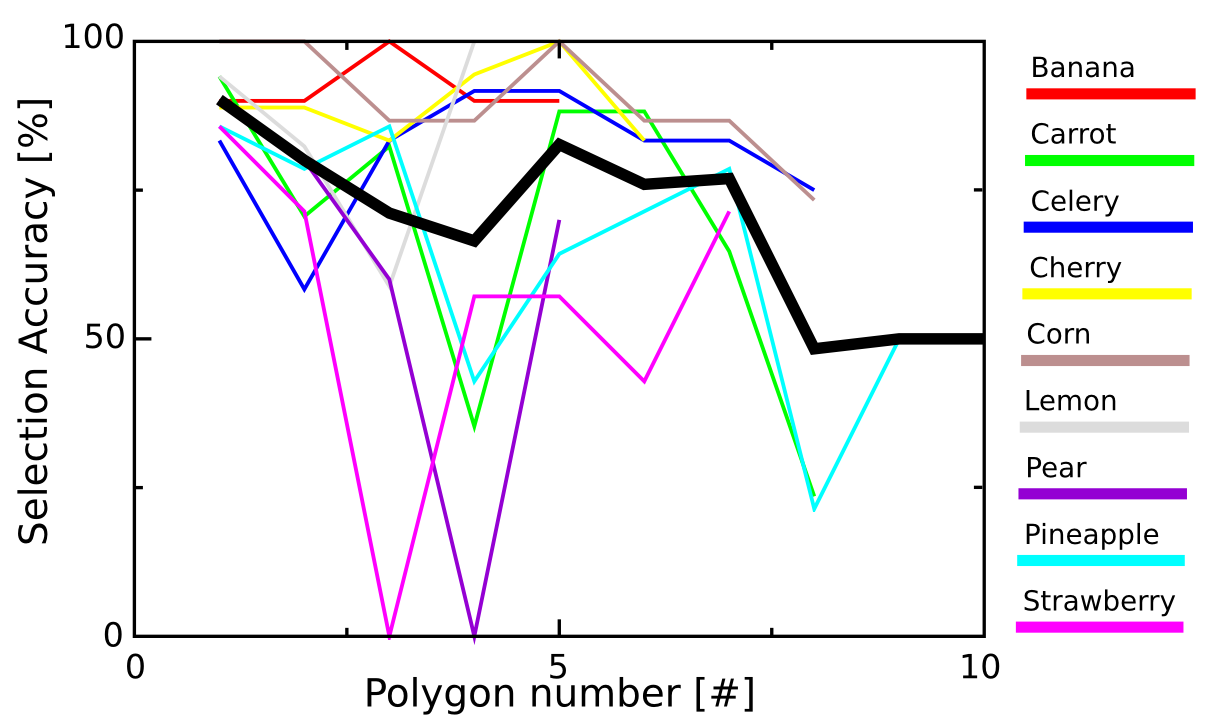}

        \caption{Performance drop in early polygons. Polygons appearing earlier in the
reconstruction task bear more visual information about the targets than those showing up later. The
former should be easier to recognize and be correctly classified more often. Despite this, there is
a marked drop in selection accuracy of intermediate polygons when accuracy is averaged over all
polygons with the same rank in the distribution (thick black line). We appreciate that the
strawberry (magenta), the carrot (green), the pear (violet), and the pineapple (cyan) pose
challenging tasks to the subjects precisely at polygons $3$ or $4$ of their decomposition,
accounting for the average drop.}

        \label{fig:8}
      \end{center}
    \end{figure}

    Figure \ref{fig:8} clarifies this drop in performance for intermediate polygons. When we look at
the average accuracy per polygon for single reconstructions we note that some of them present a
strongly marked fall in selection accuracy for polygons $3$ and $4$: the strawberry in polygon $3$;
and the carrot, the pear, and the pineapple in polygon $4$. All four cases present a selection
accuracy lower than $50\%$ for these polygons, contributing to an average performance drop across
reconstructions. These polygons happen to be just difficult to classify.

    In two of these cases (strawberry and pear) the problematic polygons are very light gray pieces
that contribute to the reconstruction (perhaps by occluding some spurious detail brought in by
earlier pieces, as with the third polygon of the cherry \cite{videoCherry}). Because they are almost
white, they are easy to miss against the blank background. In the case of the carrot the performance
drop is not so dramatic. Also, this polygon number $4$ contributes to the leaves of the carrot and,
as we saw in section \ref{sec:3.04}, that was precisely a common location of ambiguous polygons. If
some ambiguous non- targets have been overly miss-classified, some actual targets must necessarily
be missed.

    Nothing remarkable has been recognized in the case of the pineapple: the conflicting polygon
seems to be a working, non-ambiguous piece of the reconstruction and we just assume that this
precise primitive was more difficult for the subjects (note that the performance raises back to
average for polygons $6$ and $7$ of the pineapple). Among the four important deviations at
intermediate stages, this is the case with the lowest drop.

\end{document}